\documentclass[superscriptaddress,secnumarabic,twocolumn,
 amssymb,amsmath,nobibnotes,aps,prd,showkeys,showpacs,nofootinbib]{revtex4}
\usepackage{mathtools}
\usepackage{graphicx}
\usepackage{booktabs}
\usepackage{appendix}
\usepackage{color}
\usepackage[normalem]{ulem}
\usepackage[utf8]{inputenc}

\begin{document}

\title{Spinning test body orbiting around a Schwarzschild black hole: Comparing Spin Supplementary Conditions for Circular Equatorial Orbits  }
\author{Iason Timogiannis}
\email{timogian@phys.uoa.gr}
\affiliation{Section of Astrophysics, Astronomy, and Mechanics, Department of Physics,
University of Athens, Panepistimiopolis Zografos GR15783, Athens, Greece}
\author{Georgios Lukes-Gerakopoulos}
\email{gglukes@gmail.com}
\affiliation{Astronomical Institute of the Czech Academy of Sciences,
Bo\v{c}n\'{i} II 1401/1a, CZ-141 00 Prague, Czech Republic}
\author{Theocharis A. Apostolatos}
\email{thapostol@phys.uoa.gr}
\affiliation{Section of Astrophysics, Astronomy, and Mechanics, Department of Physics,
University of Athens, Panepistimiopolis Zografos GR15783, Athens, Greece}

\begin{abstract}
The Mathisson-Papapetrou-Dixon (MPD) equations  describe the motion of an extended test body in general relativity. This system of equations, though, is underdetermined and has to be accompanied by constraining supplementary conditions, even in its simplest version, which is the pole-dipole approximation corresponding to a spinning test body. In particular, imposing a spin supplementary condition (SSC) fixes the center of the mass of the spinning body, i.e.\ the centroid of the body. In the present study, we examine whether characteristic features of the centroid of a spinning test body, moving in a circular equatorial orbit around a massive black hole, are preserved under the transition to another centroid of the same physical body, governed by a different SSC. For this purpose, we establish an analytical algorithm for deriving the orbital frequency of a spinning body, moving in the background of an arbitrary, stationary, axisymmetric spacetime with reflection symmetry, for the Tulzcyjew-Dixon, the Mathisson-Pirani and the Ohashi-Kyrian-Semer\'{a}k SSCs. Then, we focus on the Schwarzschild black hole background and a power series expansion method is developed, in order to investigate the discrepancies in the orbital frequencies expanded in power series of the spin among the different SSCs. Lastly, by employing the fact that the position of the centroid and the measure of the spin alters under the centroid's transition, we impose proper corrections to the power expansion of the orbital frequencies, which allows to improve the convergence between the SSCs. Our concluding argument is that when we shift from one circular equatorial orbit to another in the Schwarzschild background, under the change of a SSC, the convergence between the SSCs holds only up to certain powers in the spin expansion, and it cannot be achieved for the whole power series. 
\end{abstract}

\pacs{~~}
\keywords{Gravitation; Dynamical systems}
\maketitle

\section{Introduction}

The motion of an extended, spinning, test body in the presence of a curved spacetime background is an interesting and relatively old problem in general relativity. Mathisson \cite{Mathisson37} and Papapetrou \cite{Papapetrou51} provided a multipole moments expansion formalism framework to address the issue, while later, in the mid 60's, Dixon \cite{Dixon64} obtained the covariant form of what is nowadays known as the Mathisson-Papapetrou-Dixon equations. In brief, the derivation of MPD equations, in the pole-dipole approximation, follows from the conservation of the stress-energy tensor $T^{\mu\nu}$ describing the body. The monopole, dipole or higher order terms (the latter is neglected in our case), are integrals of the stress-energy tensor, the integration being carried over a 3-dimensional space-like hypersurface defined by a constant, though arbitrarily chosen, coordinate time $t$. The zeroth multipole moments, often called the mass-monopole, can be encoded in the four-momentum $p^\mu$, while the first moments, often called the spin-dipole, in the antisymmetric spin tensor $S^{\mu\nu}$. The pole-dipole version of Mathisson-Papapetrou-Dixon equations is accompanied by the assumption that the extended, spinning body is a test body, hence it does not deform the background spacetime due to its own gravity and it is characterized solely by two quantities, its mass and its spin. The MPD equations in their standard form read \cite{Papapetrou51}, \cite{Dixon74b}:
\begin{align}
&\frac{Dp^\mu}{d\lambda} =-\frac{1}{2}R^\mu_{\nu\kappa\lambda} u^\nu S^{\kappa \lambda}, \label{eq:MPD_mom}\\
&\frac{DS^{\mu \nu}}{d\lambda} =p^\mu u^\nu -u^\mu p^\nu, \label{eq:MPD_spin}
\end{align}
where by definition $\displaystyle \frac{D}{d\lambda} := u^\mu \nabla_\mu$ is the covariant derivative along the four-velocity $\displaystyle u^\mu=\frac{d z^\mu (\lambda)}{d\lambda}$, where $z^\mu(\lambda)$ is the representative worldline of the test body and $\lambda$ is the proper time. For future reference, let us also introduce two different notions of the mass of the spinning body, which also ensures the timelike character of the four-momentum, that is, the dynamical rest mass $\mu:=\sqrt{-p^\nu p_\nu}$ as well as the kinematical rest mass $m:=-p^\nu u_\nu$. Furthermore, the measure of the spin is defined as:
\begin{align} \label{eq:spin_m}
 S^2=\frac{1}{2} S^{\mu\nu}S_{\mu\nu}\, ,  
\end{align}
which implies that the spin tensor is spacelike.

The notoriety of the MPD equations  is due to forming an underdefined set of equations of motion. Namely, a set of fourteen variables $\{p^\mu, u^\nu, S^{\kappa\lambda}\}$ has to be evolved with the help of Eqs.~\eqref{eq:MPD_mom}-\eqref{eq:MPD_spin} with the constraint $u^\nu u_\nu=-1$. These are three less than the number of variables. In order to overcome this obstacle, numerous constraining relations have been proposed over the years, which are known as the Spin Supplementary Conditions (SSCs). All SSCs in the existing literature share a common feature though, that they can be written in the form $V_\mu S^{\mu\nu}=0$, where $V^\mu$ is a suitable future oriented timelike four-vector, which characterizes the observer and is often normalized to $V^\nu V_\nu=-1$, just like the test body's four-velocity\footnote{Note that such a SSC imposes  exactly three linearly independent constraints.}. Fixing $V^\mu$, i.e. a SSC, fixes the center of mass (centroid) of the body, while changing a spin supplementary condition is equivalent to changing the reference vector, which physically implies that an observer tracks a different centroid for the body.

There are five established SSCs used in the literature, but in this work we focus just on three of them as done in the first paper of the series \cite{Harms16}. More specifically, we are not discussing the Corinaldesi-Papapetrou SSC \cite{Corinaldesi51} and the Newton-Wigner SSC \cite{Wigner49, Pryce48}, and we concentrate on the following ones:

\begin{itemize}
    \item The Tulczyjew-Dixon (TD) SSC \cite{Dixon70,Tulczyjew59} has as a unique reference four-vector defined as $V^\nu:=p^\nu/\mu$. Under TD SSC $S$ and $\mu$ are constants upon evolution \cite{Semerak99,LG17} independently from the background spacetime. The TD SSC is widely used in numerical calculations \cite{Mino:1995fm,Tanaka:1996ht,Suzuki:1997by,Han08,Hackmann14,Harms16,Akcay20}, mostly due to the existence of an explicit relation between the four-momentum and the four-velocity of the spinning, test body \cite{Kunzle72}:
    \begin{equation}
u^\mu=\frac{m}{\mu^2}\biggl(p^\mu+\frac{2R_{\nu\rho\kappa\lambda}p^\rho S^{\kappa\lambda}S^{\mu\nu}}{4\mu^2+R_{\alpha\beta\gamma\delta}S^{\alpha\beta}S^{\gamma\delta}}\biggr).
\end{equation}

\item The Mathisson-Pirani (MP) SSC \cite{Mathisson37, Pirani56} is defined by a timelike four-vector $V^\nu:=u^\nu$. Consequently, the observer comoves in a reference frame which coincides with the rest frame of the body. Under MP SSC  $m$ and $S$  are constants upon evolution \cite{Semerak99,LG17} independently from the background spacetime. A recent study \cite{Costa18} revealed the existence of a momentum-velocity relation for the MP SSC:
\begin{equation} \label{eq:MPmom2vel}
    u^\mu=\frac{1}{m}\biggl(p^\mu+\frac{S_{\nu\lambda}S^{\mu\nu}p^\lambda}{S^2}\biggr).
\end{equation}
However, the MP SSC are not so popular in numerical works, since the orbits produced by them tend to exhibit helices. These helices were considered for long time to be unphysical until their nature was interpreted in \cite{Costa12}. This SSC is quite popular in analytical works \cite{Costa12,Semerak15,Costa18,Plyatsko18}. 

\item The Ohashi-Kyrian-Semer\'{a}k (OKS) SSC \cite{Ohashi03,Kyrian07,Semerak15} exploits the freedom in the choice of the future-pointing timelike four-vector $V^\nu$, in order to impose a desirable feature in the MPD equations. Namely, $V^\mu$ is chosen in such way that the hidden momentum:
\begin{align} \label{eq:hidmom}
 {p_h}^\nu=u_\mu \frac{DS^{\mu \nu}}{d\lambda}
\end{align}
is eliminated \cite{Semerak15}, i.e.  ${p_h}^\nu=0$. Note that, if Eq.~\eqref{eq:MPD_spin} is contracted with $u_\mu$, one concludes that:
\begin{align}
    p^\nu= m u^\nu+ {p_h}^\nu \, .
\end{align}
Hence, OKS condition implies that $p^\nu=m u^\nu$. Under OKS SSC $\mu=m$ and $S$  are constants upon evolution \cite{Semerak99,LG17} independently from the background spacetime.
\end{itemize} 

Apart from being interesting from a theoretical point of view, the MPD equations have also an astrophysical application in modeling Extreme Mass Ratio Inspirals (EMRIs). An EMRI consists of a stellar compact object inspiralling in the background spacetime of a supermassive black hole (BH). Such physical systems are likely to exist in the center of galaxies, where stellar black holes or neutron stars, may be found orbiting around a giant, central black hole. In this work, we denote by $M$ the mass of the central, Schwarzschild BH and by $\mu$ (TD/OKS SSC), or $m$ (MP SSC) the mass of the test body, with $M \gg \mu$ and $M \gg m$, respectively. Additionally, to avoid complicated notation, the same symbol has been used for denoting the dimensionless measure of the spin $\sigma$ of the test body, that is, $\sigma:=\frac{S}{\mu M}$ under the TD and OKS SSC and $\sigma:=\frac{S}{m M}$ under MP SSC  respectively. In some parts of the paper the dimensionless spin of the test body, for the Tulczyjew-Dixon SSC, is also represented by $\tilde{\sigma}$, although the cases where that alteration takes place, will be explicitly mentioned,  in order to avoid any confusion.

The question this work addresses is: if we shift from a centroid moving on a circular equatorial orbit (CEO) in a  black hole background to another centroid of the same body, will the new centroid reproduce the characteristic physical features of the original centroid or not? To answer this question we revisit the issue of finding CEOs around a black hole in Sec.~\ref{sec:CEOs} and employ a power series expansion formalism in the majority of the steps in our work. While the technical details of this method will be presented in Sec.~\ref{sec:comp}, it is sufficient to say here, that the fundamental idea behind the power series expansion is to take advantage of the smallness of the dimensionless spin $\sigma$ of the inspiralling body. In the case of an EMRI,  when the extended, spinning, test body represents a rotating black hole or a neutron star, then  $\lvert\sigma\rvert\lesssim \frac{\mu}{M} \ll 1$ (for TD and OKS SSCs) and $\lvert\sigma\rvert\lesssim \frac{m}{M}\ll 1$ (for MP SSC) \cite{Hartl03}.  

\textit{Units and notation:} All calculations have been made in geometric units, in which the speed of light and the gravitational constant are set to  $c=G=1$. Moreover, the Riemann tensor is defined as $R^\mu_{\nu\kappa\lambda}=\Gamma^\mu _{\kappa \alpha} \Gamma^\alpha _{\lambda \nu}-\partial_\lambda \Gamma^\mu _{\kappa \nu}--\Gamma^\mu _{\lambda\alpha}\Gamma^\alpha _{\kappa \nu}+\partial_\kappa \Gamma^\mu _{\lambda \nu}$, whereas the Christoffel symbols are computed from the metric with signature $(-,+,+,+)$, expressed in terms of the standard Boyer-Lindquist coordinates $\{t,r,\theta,\phi\}$. Einstein's summation convention has been followed, with all indices running from 0 to 3. The Levi-Civita tensor is given by $\epsilon_{\mu\nu\rho\sigma}=\sqrt{-g}\tilde{\epsilon}_{\mu\nu\rho\sigma}$, with the Levi-Civita symbol $\tilde{\epsilon}_{t r \theta \phi}=1$ and $g$ the determinant of the background metric.

\section{Circular Equatorial Orbits} \label{sec:CEOs}

The problem of finding CEOs for the MPD equations reduces to the selection of appropriate initial data for the variables $\{z^\nu, p^\nu,  S^{\mu\nu}\}$, so that circular equatorial motion is obtained upon evolution of the equations. Both numerical and semianalytic methods, based on the concept of effective potentials \cite{Tod76,Steinhoff12,Hackmann14,Harms16}, have been in development since the pioneering work of Rasband \cite{Rasband73}. Khodagholizadeh et al. \cite{Khodagholizadeh20}, on the other hand, followed a more straightforward and purely analytical approach, by evaluating all the non vanishing components of the MPD equations for a Kerr background under any SSC, but obtained results for only the TD and the MP conditions. In this section, we generalize this analytical procedure for arbitrary, stationary, axisymmetric spacetimes, with reflection symmetry (SAR spacetimes); we then focus on TD, MP and OKS SSCs and use Schwarzschild spacetime as a specific example for each SSC.   

In BL coordinates the line element of a SAR spacetime reads:
\begin{align*}
    ds^2=g_{tt} dt^2+2 g_{t\phi} dt d\phi+g_{\phi\phi}d\phi^2+g_{rr} dr^2+g_{\theta\theta} d\theta^2\, .
\end{align*}
The metric components $g_{\mu\nu}$ are just functions of $r$ and $\theta$. On the equatorial plane, due to the reflection symmetry, it holds that $\partial_\theta g_{\mu\nu}=0$. In the specific case of the Schwarzschild spacetime, on which we focus later on, the metric components reduce to:
\begin{align*}
    g_{tt}&=-g_{rr}^{-1}=-\left(1-\frac{2 M}{r}\right)\, ,\quad g_{t\phi}=0\, , \\ g_{\theta\theta}&=r^2\, , \quad g_{\phi\phi}=r^2 \sin^2 \theta\, .
\end{align*}

Without loss of generality, we identify the time coordinate of the test body with the background coordinate time $t$. For a CEO the spatial coordinates are $r=constant$, $\theta=\frac{\pi}{2}$ and $\phi=\Omega t$, where $\Omega=u^\phi/u^t$ is the orbital frequency of the spinning, test body; the radial and polar four-velocity components has to be $u^r=0$ and $u^\theta=0$ respectively, while the constraints $p^r=0$ and $p^\theta=0$ hold for all the SSCs examined in our work \cite{Harms16}. The normalization condition of the four-velocity implies that:
\begin{equation} \label{eq:vel_t}
u^t=\frac{1}{\sqrt{-g_{tt}-2g_{t\phi}\Omega-g_{\phi\phi}\Omega^2}}.
\end{equation}
Moreover, the demand that the spin four-vector:
\begin{align}
\label{eq:SpinVect}
 S_\mu := -\frac{1}{2} \epsilon_{\mu\nu\rho\sigma}
          \, V^\nu \, S^{\rho\sigma} \; ,
\end{align}
is aligned (or anti-aligned) with the component of the total angular momentum $J_z$ that is perpendicular ($z$-axis) to the equatorial plane\footnote{Note that in the case of the TD and MP SSCs we do not have to actually demand the alignment, since it is a natural outcome of the $u^r=u^\theta=0$  restriction \cite{Steinhoff12,Harms16}.}, $S^\nu=S^\theta \delta^\nu_\theta$,  leaves the spin tensor only with the following nonzero components:
\begin{align} 
    S^{tr}&=-S^{rt}=-S\sqrt{-\frac{g_{\theta\theta}}{g}}V_\phi, \label{eq:s1}\\
    S^{r\phi}&=-S^{\phi r}=-S\sqrt{-\frac{g_{\theta\theta}}{g}}V_t, \label{eq:s2}
\end{align}
where the inverse relation of Eq.~\eqref{eq:SpinVect}:
\begin{align}
    S^{\rho\sigma}=-\epsilon^{\rho\sigma\nu\kappa}S_\nu V_\kappa,
\end{align}
and the fact that $S=\sqrt{g_{\theta\theta}}S^{\theta}$ have been taken into account. 

SAR spacetimes are characterized by two Killing vector fields, a time-like one $\xi^\mu_{(t)}$ and a space-like one $\xi^\mu_{(\phi)}$. The former field leads to the conservation of the energy $E$, while the latter to the conservation of the $z$ component of the total angular momentum $J_z$ of a spinning, test body moving in a SAR spacetime. On a CEO these constants of motion read \cite{Harms16}:
\begin{align}
    &E=-p_t+\frac{S}{2}\sqrt{-\frac{g_{\theta\theta}}{g}}\biggl(g_{t\phi,r}V_t-g_{tt,r}V_\phi\biggr), \label{eq:energy} \\ 
    &J_z=p_\phi+\frac{S}{2}\sqrt{-\frac{g_{\theta\theta}}{g}}\biggl(g_{t\phi,r}V_\phi-g_{\phi\phi,r}V_t\biggr). \label{eq:angmom}
\end{align}

For CEOs the set of the MPD equations~\eqref{eq:MPD_mom}, \eqref{eq:MPD_spin}, leads to trivial identities, apart from the case of the $\frac{Dp^r}{d\lambda}$ and $\frac{DS^{t\phi}}{d\lambda}$ components. For those components it is true that  $\frac{d p^r}{d\lambda}=\frac{dS^{t\phi}}{d\lambda}=0$ and after some calculations one concludes that:
\begin{align}
    &R^r_{\nu\kappa\lambda}u^\nu S^{\kappa\lambda}=-2 \Gamma^r_{\pi\sigma}u^\pi p^\sigma, \label{eq:MPD1} \\
    &p^t u^\phi-p^\phi u^t=\Gamma ^t_{\kappa r}u^\kappa S^{r \phi}+\Gamma^\phi_{\sigma r}u^\sigma S^{t r}. \label{eq:MPD2}
\end{align}
As a result, for the special case of CEOs on SAR spacetimes, these two equations are satisfied simultaneously. In the rest of the section, we describe the techniques employed to find the orbital frequency of the spinning body, for each spin supplementary condition separately, based only on the relations~\eqref{eq:MPD1} and \eqref{eq:MPD2}. 
    
\subsection{Tulzcyjew-Dixon SSC}

For the TD SSC the reference four-vector has to be replaced by $ V^\nu=p^\nu/\mu$ in Eqs.~\eqref{eq:s1} and \eqref{eq:s2}. The spin tensor components are, then, plugged into the system~\eqref{eq:MPD1}, \eqref{eq:MPD2}, which can be rearranged in order to give $\frac{p^t}{p^\phi}$ in terms of the orbital frequency of the test body as well as the geometric characteristics of the spacetime itself. In particular, one gets:
\begin{align}
        &\frac{p^t}{p^\phi}=f_1(r,\Omega;S), \label{eq:TD1}\\
        &\frac{p^t}{p^\phi}=f_2(r,\Omega;S), \label{eq:TD2}
\end{align}
where:
\begin{widetext}
\begin{align} \label{eq:appfirst}
&f_1(r,\Omega;S)=\frac{-\mu \biggl(\Gamma^r_{t\phi}+ \Omega \Gamma^r_{\phi\phi}\biggr)+S \sqrt{-\frac{g_{\theta\theta}}{g}}
    \biggl[g_{\phi\phi}\biggl(R^r_{ttr}+\Omega R^r_{\phi t r}\biggr)-g_{t\phi}\biggl(R^r_{\phi t r}-\Omega R^r_{\phi r \phi}\biggr)\biggr]}{\mu \biggl(\Gamma^r_{tt}+\Omega \Gamma^r_{t\phi}\biggr)-S\sqrt{-\frac{g_{\theta\theta}}{g}}
     \biggl[g_{t\phi}\biggl(R^r_{ttr}+\Omega R^r_{\phi t r}\biggr)-g_{tt}\biggl(R^r_{\phi tr}-\Omega R^r_{\phi r \phi}\biggr)\biggr]},\\ \label{eq:appsecond}
     &f_2(r,\Omega;S)=\frac{\mu-S\sqrt{-\frac{g_{\theta\theta}}{g}}\biggl[g_{\phi\phi}\biggl(\Gamma^\phi_{tr}
    +\Omega\Gamma^\phi_{\phi r}\biggr)+g_{t\phi}\biggl(\Gamma^t_{tr}+\Omega\Gamma^t_{\phi r}\biggr)\biggr]}{\mu\Omega+S\sqrt{-\frac{g_{\theta\theta}}{g}}\biggl[g_{t\phi}\biggl(\Gamma^\phi_{tr}+\Omega\Gamma^\phi_{\phi r}\biggr)+g_{tt}\biggl(\Gamma^t_{tr}+\Omega\Gamma^t_{\phi r}\biggr)\biggr]},
\end{align}
\end{widetext}  
are irreducible fractions. These relations are valid for every SAR spacetime. By equating the right-hand sides of Eqs.~\eqref{eq:TD1}, \eqref{eq:TD2}, we obtain a quadratic equation for $\Omega$, with typically, two physically accepted solutions, corresponding to corotation $\Omega_+$ and counterrotation $\Omega_-$, reading:
\begin{equation} \label{eq:2vathmia}
    \Omega_{\pm}=\frac{-\rho_2\pm \sqrt{\rho_2^2-4\rho_1\rho_3}}{2\rho_1}\, ,
\end{equation}
where the lengthy expressions $\rho_1$, $\rho_2$, $\rho_3$  are given in Appendix~\ref{sec:app1}.

Note that, in order to obtain $p^t$ and $p^\phi$ as well, one has to employ the definition of the dynamical rest mass $\mu:=\sqrt{-p^\nu p_\nu}$, which leads to:

 \begin{align} 
        &p^t=\pm \frac{\mu f_1}{\sqrt{-g_{tt}f_1^2-2g_{t\phi}f_1-g_{\phi\phi}}}, \label{eq:pt_TD}\\
        &p^\phi=\pm \frac{\mu}{\sqrt{-g_{tt}f_1^2-2g_{t\phi}f_1-g_{\phi\phi}}}, \label{eq:pf_TD}
    \end{align}
where the plus sign corresponds to $\Omega_+$, while the minus sign is related to $\Omega_-$. We wish to underline at this point that, the ``positive'' orbital frequencies $\Omega_{+}$ are associated with $E>0$ and $J_z>0$, while the ``negative'' orbital frequencies $\Omega_{-}$ are associated with $E>0$ and $J_z<0$. This is a rather universal pattern, which characterizes all SSCs examined here. Eqs.~\eqref{eq:pt_TD},~\eqref{eq:pf_TD} will be needed in our analysis in Sec.~\ref{sec:comp}.     

The quadratic equation for the orbital frequency of an extended, spinning, test body, in the Schwarzschild black hole limit is reduced to:
\begin{equation}\label{eq:SchCEOfr_TD}
    r^3\biggl(r^3 \mu^2-M S^2\biggr)\Omega^2+3M \mu\Omega S r^3-M\biggl(r^3 \mu^2+2M S^2\biggr)=0,
\end{equation}
which coincides with the expression presented in \cite{Khodagholizadeh20}, when the Kerr parameter is set to zero. The solutions of such an equation and the behaviour of its discriminant, have been thoroughly examined by numerous authors and in more general contexts ($a\neq0)$. Consequently, the details regarding the $\Omega_{\pm}$ roots have been omitted and we will proceed with the investigation of the Mathisson-Pirani SSC.
    
\subsection{Mathisson-Pirani SSC}

Imposing the Mathisson-Pirani condition implies that $V^\nu=u^\nu$. The definition equation of the kinematical rest mass, $m:=-p^\nu u_\nu$, provides here, a relation, which when combined with Eq.~\eqref{eq:MPD2} can be reduced to:
\begin{align}
  &p^t=mu^t+h_1(r,\Omega;S), \label{eq:MP1}\\
  &p^\phi=mu^\phi-h_2(r,\Omega;S), \label{eq:MP2}
\end{align}
where the functions:
\begin{widetext}
\begin{align*} 
&h_1(r,\Omega;S)=S \sqrt{-\frac{g_{\theta\theta}}{g}}
\frac{g_{t\phi} +g_{\phi\phi}\Omega}
{\left(-g_{tt}-2g_{t\phi}\Omega-g_{\phi\phi}\Omega^2\right)^{3/2}}
\biggl[\biggl(g_{tt} +g_{t\phi} \Omega\biggr)\biggl(\Gamma^t_{tr} +\Omega\Gamma^t_{\phi r} \biggr)+\biggl(g_{t\phi} +g_{\phi\phi} \Omega\biggr)\biggl(\Gamma^\phi_{tr} +\Omega\Gamma^\phi_{\phi r} \biggr)\biggr],\\
&h_2(r,\Omega;S)=S \sqrt{-\frac{g_{\theta\theta}}{g}}
\frac{g_{tt} +g_{t\phi} \Omega}
{\left(-g_{tt}-2g_{t\phi}\Omega-g_{\phi\phi}\Omega^2\right)^{3/2}}
\biggl[\biggl(g_{tt} +g_{t\phi} \Omega\biggr)\biggl(\Gamma^t_{tr} +\Omega\Gamma^t_{\phi r} \biggr)+\biggl(g_{t\phi} +g_{\phi\phi} \Omega\biggr)\biggl(\Gamma^\phi_{tr} +\Omega\Gamma^\phi_{\phi r} \biggr)\biggr],
\end{align*}
\end{widetext}
represent the hidden momentum\footnote{This hidden momentum may be cast as analogous, albeit with a very different nature, to the hidden momentum of electromagnetic systems.} of the test body for CEOs, i.e. a correction made to the relation between the four-momentum and the four-velocity, due to the presence of the spin. 

Inserting the constraints~\eqref{eq:MP1},~\eqref{eq:MP2} into Eq.~\eqref{eq:MPD1} yields a quartic equation for the orbital frequency and more precisely:
\begin{equation} \label{eq:4vathmia}
 \xi_1\Omega^4+\xi_2\Omega^3+\xi_3\Omega^2+\xi_4\Omega+\xi_5=0, \end{equation}
with the polynomial coefficients $\xi_i$, presented in the Appendix \ref{sec:app1}. 

Such an equation, which is analytically solvable using certain algebraic techniques, leads in general to four distinct solutions, two of them are considered to be unnatural \cite{Costa18}, while the remaining two pertain to $\Omega_+$ and $\Omega_-$. The procedure followed in order to reject the unphysical solutions for the orbital frequency, strongly depends on the metric used to describe spacetime and as a result will be thoroughly discussed in the next paragraph, where we consider the simple example of the Schwarzschild spacetime. Once the frequency $\Omega$ has been determined, the energy and the total angular momentum along the z-axis for the test body, follows from Eqs.~\eqref{eq:energy}, \eqref{eq:angmom} for $V^\nu=u^\nu$, when the relation $u^\phi=\Omega u^t$ and Eq.~\eqref{eq:vel_t} are taken into account.
    
As a simple crosscheck of the lengthy expressions for $h_{1,2}$, note that, the hidden momentum components for the Schwarzschild case can be written:
\begin{equation*}
 h_{1,2}(r,\Omega;S)\propto S\biggl(r^3\Omega^2-M\biggr), 
\end{equation*}
that, as one expects, reduces to zero, in the geodesic limit, where the spin vanishes and $\Omega$ acquires its Keplerian value. The aforementioned quartic equation with respect to the orbital frequency for the Schwarzschild spacetime reads:
\begin{align} \label{eq:SchCEOfr_MP}
    &mr^6\Omega^4-Sr^3\biggl(r-6M\biggr)\Omega^3-mr^3\biggl(r-M\biggr)\Omega^2 \nonumber\\ 
    &-M\Omega S\biggl(2r-3M\biggr)+mM\biggl(r-2M\biggr)=0. 
\end{align}
As a second crosscheck, record that, Eq. \eqref{eq:SchCEOfr_MP} is identical to the one derived in \cite{Khodagholizadeh20}, when $a\to 0$. Additionally, the physically acceptable solutions (corotating and counterrotating) are chosen so that $\Omega_{\pm}\propto \pm r^{-\frac{3}{2}}$, for a non-spinning body. An extended discussion concerning the generic roots of polynomial~\eqref{eq:SchCEOfr_MP} can be found in \cite{Costa18}, while the explicit expressions of the roots can be found in \cite{Costa18} and a Taylor expansion in \cite{Khodagholizadeh20}.

\subsection{Ohashi-Kyrian-Semer\'{a}k SSC}
    
The algorithm used for acquiring solutions for CEOs, under the OKS condition, had been until recently an uncharted territory. In \cite{Harms16}, it has been shown that such an algorithm exists for SAR spacetimes by employing a system of three effective potentials. In the present study we introduce a novel method of finding CEOs, by combining Eqs.~\eqref{eq:MPD1} and \eqref{eq:MPD2} with the normalization conditions of $u^\nu$ and $V^\nu$ and the fact that for OKS $p^\mu=m u^\mu$.

For the OKS SSC we leave the denotation of the reference four-vector, in its general form, i.e. $V^\nu$. The relation $V^\nu V_\nu=-1$ is in fact, a quadratic equation for $V^t$, with solution:
\begin{equation} \label{eq:vt}
 V^t=-\frac{g_{t\phi}V^\phi+\sqrt{(g_{t\phi}V^\phi)^2-g_{tt}[g_{\phi\phi}(V^\phi)^2+1]}}{g_{tt}},
\end{equation}
where the $-$ sign has been chosen so that $V^t>0$ in the Schwarzschild black hole limit. The latter claim is justified by the fact that the time coordinate for the extended, test body is identified with the background coordinate time. 

Eqs.~\eqref{eq:MPD2}, \eqref{eq:vt}, combined with the normalization condition of the four-velocity and the relation $u^\phi=\Omega u^t$, reveal $V^\phi$ in terms of the orbital frequency:
    \begin{widetext}
    \begin{equation} \label{eq:OKS1}
        V^\phi=\pm \sqrt{-\frac{g_{tt}\biggl[g_{tt}\biggl(\Gamma^t_{tr}+\Omega\Gamma^t_{\phi r}\biggr)+g_{t\phi}\biggl(\Gamma^\phi_{tr}+\Omega\Gamma^\phi_{\phi r}\biggr)\biggr]^2}{\biggl(g_{tt}g_{\phi\phi}-g_{t\phi}^2\biggr)\biggl\{\biggl(g_{tt}g_{\phi\phi}-g_{t\phi}^2\biggr)\biggl(\Gamma^\phi_{tr}+\Omega\Gamma^\phi_{\phi r}\biggr)^2+\biggl[g_{tt}\biggl(\Gamma^t_{tr}+\Omega\Gamma^t_{\phi r}\biggr)+g_{t\phi}\biggl(\Gamma^\phi_{tr}+\Omega\Gamma^\phi_{\phi r}\biggr)\biggr]^2\biggr\}}}.
    \end{equation}
    \end{widetext}

Since $V^t$, $V^\phi$, along with the four-velocity components have been expressed in terms of the orbital frequency, Eq.~\eqref{eq:MPD1}  transforms into a twelfth degree polynomial in $\Omega$, after various, complex manipulations. It is crucial to notice here that, both choices for the sign of $V^\phi$ lead to the same equation, which can be solved analytically, at least for the Schwarzschild case, where it reduces to a sextic equation. However, the general expressions for the several polynomial coefficients are rather extended and will not be provided in the present work. We would prefer instead, to present a middle step of the computation, for the shake of completeness. Thus, by defining the parameters:
\begin{align*}
& K=\Gamma^r_{tt}+2\Omega \Gamma^r_{t\phi}+\Omega^2\Gamma^r_{\phi\phi},\\
 &\Lambda=g_{t\phi}R^r_{ttr}+g_{tt}R^r_{tr\phi}+\Omega\biggl(g_{tt}R^r_{\phi r \phi}-g_{t\phi}R^r_{tr\phi}\biggr), 
 \end{align*}
the following expression is derived:
\begin{widetext}
\begin{align}
    &\biggl[2mgKS \sqrt{-\frac{g_{\theta\theta}}{g}}g_{tt}\biggl(g_{t\phi}^2-g_{tt}g_{\phi\phi}\biggr)\biggl(R^r_{ttr}-\Omega R^r_{tr\phi}\biggr)u^tV^\phi\biggr]^2=\biggl\{-\Lambda^2S^2 g_{\theta\theta} \biggl[\biggl(g_{t\phi}^2-g_{tt}g_{\phi\phi}\biggr)\biggl(V^\phi\biggr)^2-g_{tt}\biggr]\nonumber\\
    &-g\biggl(m K g_{tt}u^t\biggr)^2+S^2 g_{\theta\theta}\biggl(g_{t\phi}^2-g_{tt}g_{\phi\phi}\biggr)^2\biggl(R^r_{ttr}-\Omega R^r_{tr\phi}\biggr)^2\biggl(V^\phi\biggr)^2\biggr\}^2.
\end{align}
\end{widetext}

As it is not trivial to generally solve the twelfth degree polynomial equation, satisfied by the test body's orbital frequency, we will employ the proposed, innovative procedure in the special case of a central Schwarzschild black hole. Under this restriction, Eq.~\eqref{eq:MPD1} yields:  
\begin{equation} \label{eq:app4}
    m u^t\biggl(M-r^3\Omega^2\biggr)=MS\biggl(2V^\phi+\Omega V^t\biggr),
    \end{equation}
    while Eq.~\eqref{eq:MPD2} can be written in the form:
    \begin{equation} \label{eq:app5}
    r^3\Omega V^\phi  =M V^t.
    \end{equation}
    The system of Eqs.~\eqref{eq:app4}, \eqref{eq:app5} is supplemented by an auxiliary correlation, that is, the normalization condition of the reference four-vector $V^\nu V_\nu=-1$, along with Eq.~\eqref{eq:vel_t}. Under these constraints, we arrive at a sixth order polynomial reading: 
\begin{widetext}
\begin{align} \label{eq:SchCEOfr_OKS}
     &r^9\biggl[m^2\biggl(r-2M\biggr)r^3+M^2 S^2\biggr]\Omega^6-M r^6\biggl[m^2\biggl(2r-3M\biggr)r^3
    +M S^2\biggl(r-6M\biggr)\biggr]\Omega^4\nonumber \\
    &+M^2 r^3\biggl[m^2 r^4-4M S^2\biggl(r-3M\biggr)\biggr]\Omega^2
    -M^4\biggl[m^2 r^3+4S^2\biggl(r-2M\biggr)\biggr]=0.
\end{align}
\end{widetext}
It is crucial to observe at this point that, Eq.~\eqref{eq:SchCEOfr_OKS} is in fact a cubic equation in $\Omega^2$, with rigorously computable, analytical roots. The resulting expressions however, are extensive and cannot be presented here. We will instead focus our attention on the method of choosing the natural solutions. First and foremost, the roots that do not approach the Keplerian frequency, when the spin vanishes, have to be rejected. Additionally, the remaining solutions for the orbital frequency, which satisfy the original Eqs. \eqref{eq:app4}, \eqref{eq:app5} as well, form a branch, corresponding to corotation and counterrotation, respectively. We have tested the proposed technique, by reproducing the values in TABLE III of \cite{Harms16}. More importantly, the entries closer to the horizon, denoted by backslash in the aforementioned paper, have also been determined with our method.
      
\section{Comparisons} \label{sec:comp}

\begin{table*}[ht] 
\centering
   \renewcommand{\arraystretch}{1.4}
  \large
   \begin{tabular}{ |c| c| c| c|}
\hline

$ \hat{\Omega}_n $ & TD SSC & MP SSC &  OKS SSC \\   \hline

$\mathcal{O}(\sigma^0)$ & $\frac{1}{\sqrt{\hat{r}^3}}$& $\frac{1}{\sqrt{\hat{r}^3}}$ & $\frac{1}{\sqrt{\hat{r}^3}}$ \\ \hline
$\mathcal{O}(\sigma^1)$ &$-\frac{3}{2\hat{r}^3}$ & $-\frac{3}{2\hat{r}^3}$ &$-\frac{3}{2\hat{r}^3}$ \\ \hline
$\mathcal{O}(\sigma^2)$&$\frac{21}{8\sqrt{\hat{r}^9}}$& $\frac{21}{8\sqrt{\hat{r}^9}}$&$-\frac{3(5\hat{r}-3)}{8(\hat{r}-3)\sqrt{\hat{r}^9}}$ \\ \hline
$\mathcal{O}(\sigma^3)$&$-\frac{3}{2\hat{r}^6}$& $-\frac{3(4\hat{r}-9)}{2(\hat{r}-3)\hat{r}^6}$&$-\frac{3(8\hat{r}^2-21\hat{r}+18)}{4(\hat{r}-3)^2\hat{r}^6}$ \\ \hline
$\mathcal{O}(\sigma^4)$&$\frac{39}{128\sqrt{\hat{r}^{15}}}$& $\frac{3(685\hat{r}^2-2670\hat{r}+2421)}{128(\hat{r}-3)^2\sqrt{\hat{r}^{15}}}$&$-\frac{3(971\hat{r}^3-3915\hat{r}^2+5625\hat{r}-2673)}{128(\hat{r}-3)^3\sqrt{\hat{r}^{15}}}$ \\ \hline
\end{tabular}
   \caption[caption]{The table presents the power series expansion coefficients of the dimensionless, positive orbital frequency $\hat{\Omega}_+$ of an arbitrary CEO at radius $\hat{r}$, for the three SSCs: TD, MP and OKS.}
   \label{tab:CEOFrp}
\end{table*}

In this section, we implement the methods for finding CEOs in the Schwarzschild spacetime and investigate the connection between the three different SSCs. The latter is achieved by comparing the orbital frequency for CEOs expanded in powers of the  test body's spin, with respect to their dependence on the spin itself. The orbital frequency in the corresponding polynomial equations of Sec.~\ref{sec:CEOs}, will be replaced by $\hat{\Omega}=\hat{\Omega}_n \sigma^n+\mathcal{O}\left(\sigma^5\right)$, with $n=0,1,2,3,4$, where by definition $\hat{\Omega}:=M \Omega$ is the dimensionless analogue of $\Omega$. In practice, in the perturbative calculations examined here, both notions of mass as well as $M$ scale away, so that we work with the dimensionless $\hat{r}:=r/M$, $\hat{\Omega}$ and $\sigma$. 

Before we proceed, let us also note that in the case of the Schwarzschild spacetime, due to the spherical symmetry, it is true that $\hat{\Omega}_+(-\sigma)=-\hat{\Omega}_-(\sigma)$, which implies that the alteration in the sign of the terms $\hat{\Omega}_n$ of the positive and negative orbital frequencies, only occurs at the even terms. To avoid redundancy we present only the $\Omega_{+}$ branches, but all the obtained results are the same for the $\Omega_{-}$ branches.

The value of the coefficient $\hat{\Omega}_0$ is derived by solving the respective polynomial equations~\eqref{eq:SchCEOfr_TD},~\eqref{eq:SchCEOfr_MP},~\eqref{eq:SchCEOfr_OKS} for each SSC in  the geodesic limit $\sigma\rightarrow 0$. All of the $\hat{\Omega}_0$, at least the physically meaningful ones, coincide with the Keplerian  frequency. The higher order contributions to the frequency expansion are obtained, once again, by the same polynomial equations~\eqref{eq:SchCEOfr_TD},~\eqref{eq:SchCEOfr_MP},~\eqref{eq:SchCEOfr_OKS} for every spin order term $\hat{\Omega}_n$ separately. We shall mention here that, the expansion has been terminated at $n=4$, since up to that order the discrepancies between the three SSCs have already shown up. Also, it is worth noticing that the power series expansion technique can be employed on any non linear algebraic equation, especially when radicals are involved in its roots. Thus, instead of analytically solving Eqs.~\eqref{eq:SchCEOfr_TD},~\eqref{eq:SchCEOfr_MP},~\eqref{eq:SchCEOfr_OKS}, one just has to satisfy five, admittedly simpler, linear equations, in order to determine $\hat{\Omega}_n$, for every SSC separately. 

The results are summarized in Table~\ref{tab:CEOFrp}, where a known, general claim is recovered, that is, all the SSCs are equivalent up to the linear approximation in spin\footnote{This claim is not completely accurate, the discussion below Eq.~\eqref{eq:shift1} provides some clues regarding this, but we would not deal with this subtle issue in this work.}, i.e. all the $\hat{\Omega}_1$ terms are the same. On the other hand, the power series expansion method indicates that the TD and the MP SSCs for CEOs appear to have an up to $\sigma^2$-order convergence. Let us now see what happens for a special CEO, known as the innermost stable circular orbit (ISCO).  

\subsection{ISCO Orbital Frequency}

\begin{table}[h]

  \large
   \renewcommand{\arraystretch}{1.3}
   \centering
   \begin{tabular}{ | c | c| c| c|  }
\hline

$ \hat{r}_n^{\rm ISCO} $ & TD SSC & MP SSC &  OKS SSC \\   \hline

$\mathcal{O}(\sigma^0)$ & $6$& $6$ & $6$ \\ \hline
$\mathcal{O}(\sigma^1)$ &$-\frac{2\sqrt{6}}{3}$ & $-\frac{2\sqrt{6}}{3}$ &$-\frac{2\sqrt{6}}{3}$ \\ \hline
$\mathcal{O}(\sigma^2)$&$-\frac{29}{72}$& $-\frac{29}{72}$&$-\frac{37}{72}$ \\ \hline
$\mathcal{O}(\sigma^3)$&$-\frac{137}{432\sqrt{6}}$& $-\frac{145}{432\sqrt{6}}$&$-\frac{537}{432\sqrt{6}}$ \\ \hline
$\mathcal{O}(\sigma^4)$&$-\frac{1497}{31104}$& $-\frac{2593}{31104}$&$-\frac{19505}{31104}$ \\ \hline
     \end{tabular}
     \caption[caption]{The panel depicts the power series expansion coefficients of dimensionless ISCO radius for the TD, MP and OKS SSCs.}
     \label{tab:ISCOr}
\end{table}

\begin{table}[h]
   \renewcommand{\arraystretch}{1.3}
   \centering
   \large
   \begin{tabular}{ | c | c| c| c|  }
\hline

$ \hat{\Omega}_n^{\rm ISCO}$ & TD SSC & MP SSC &  OKS SSC \\   \hline

$\mathcal{O}(\sigma^0)$ & $\frac{1}{6\sqrt{6}}$& $\frac{1}{6\sqrt{6}}$ & $\frac{1}{6\sqrt{6}}$ \\ \hline
$\mathcal{O}(\sigma^1)$ &$\frac{1}{48}$ & $\frac{1}{48}$ &$\frac{1}{48}$ \\ \hline
$\mathcal{O}(\sigma^2)$&$\frac{97}{3456\sqrt{6}}$& $\frac{97}{3456\sqrt{6}}$&$\frac{97}{3456\sqrt{6}}$ \\ \hline
$\mathcal{O}(\sigma^3)$&$\frac{1187}{186624}$& $\frac{1187}{186624}$&$\frac{2021}{186624}$ \\ \hline
$\mathcal{O}(\sigma^4)$&$\frac{105757}{11943936\sqrt{6}}$& $\frac{119645}{11943936\sqrt{6}}$&$\frac{392637}{11943936\sqrt{6}}$ \\ \hline
     \end{tabular}
     \caption[caption]{The power series expansion coefficients for the $\hat{\Omega}_+$ at ISCO around a Schwarzschild black hole, for the three SSCs (TD,MP,OKS).}
 \label{tab:ISCOFrp}    
\end{table}

An ISCO is the closest to a black hole stable CEO\footnote{Actually an ISCO is an indifferently stable orbit, see the discussion in \cite{Harms16}.}. Depending on the spin's sign the ISCO radius shifts towards or away from the horizon \cite{Harms16}. The principal aim of this subsection is to derive fourth order polynomial functions with respect to $\sigma$ of the orbital frequency computed at ISCO for each SSC. The simplest way to achieve this is to substitute the radius of a generic CEO in Table \ref{tab:CEOFrp}, by its corresponding ISCO power series expansion. The power series of the ISCO radius for the three examined SSCs is presented in Table~\ref{tab:ISCOr}. These series have been obtained by employing the effective potentials method for each SSC. The  interested reader is referred to \cite{Harms16,Hackmann14} on how to use this method.

The results are listed in Table~\ref{tab:ISCOFrp}, where the power series of $\hat{\Omega}_+$ shows that, by restricting the comparison on ISCO the differences in the $\sigma^3$-order term between the TD and the MP SSC get fixed, and the $\sigma^2$-order term between the TD,~MP SSCs and the OKS SSC become equivalent. This indicates that by shifting properly the centroids' positions, the convergence of the power series expansions between SSCs could improve as well. 

\section{Centroids' Corrections} \label{sec:Corr}

One potential way of explaining the observed differences among the aforementioned SSCs is by noting that the change of the SSC is physically equivalent to altering the representative centroid with respect to which the moments are evaluated. Whereas the contravariant components of the four-momentum do not depend on the choice of that centroid, the spin tensor has to be adjusted as follows:
\begin{equation} \label{eq:shift1}
    \tilde{S}^{\mu\nu}=S^{\mu\nu}+p^\mu \delta z^\nu-p^\nu \delta z^\mu,
\end{equation}
when the centroid is shifted from $z^\nu$ to $\tilde{z}^\nu$ due to the change of the SSC. We will denote the quantities calculated at the TD centroid with a tilde over them from now on. The quantities that do not have a tilde over them will refer to the MP or the OKS frame; each time  we will mention which case is which. In a few words, we shift from a MP/OKS SSC frame to the TD frame so that all cases are compared on equal footing. This choice is purposeful, since from the three SSCs only the TD centroid is uniquely determined, the other two hide an extra degree of freedom \cite{Costa18,Witzany19}. In the case of CEOs this extra degree of freedom is fixed for the MP and OKS SSCs, i.e. we know the four-acceleration for the MP SSC and the reference vector of the OKS SSC \cite{Harms16,Costa18}. Hence, there is no need to investigate these extra degrees of freedom which could have been induced by a centroid shift or deal with the related consequences. For instance, if we had decided to shift from a TD/OKS SSC to a MP SSC, then we would have to overcome the issue of whether the MP centroid follows a helical motion or not \cite{Costa18}.      

If we contract Eq.~\eqref{eq:shift1} with $\tilde{p}_\mu$ and by assuming \cite{Kyrian07} that:
\begin{align}\label{eq:shit_con}
  \tilde{p}_\mu \delta z^\mu=0\, ,  
\end{align}
we arrive at:
\begin{equation} \label{eq:shift2}
    \delta z^\nu=\frac{\tilde{p}_\mu S^{\mu\nu}}{\tilde{\mu}^2},
\end{equation}
where by definition, $\tilde{\mu}^2=-\tilde{g}_{\kappa\sigma}p^\kappa p^\sigma$ is the dynamical rest mass. The choice of the constraint~\eqref{eq:shit_con} is in detail discussed in Appendix~\ref{sec:shifts} along with other choices; at this point we just state that it is  one of the few viable ones and when it is combined with the assumptions made in Sec.~\ref{sec:CEOs}, concerning CEOs, implies that the position of the new centroid is shifted along the radial direction. In other words, $\delta r $ is the only non-vanishing component of Eq.~\eqref{eq:shift2}. 

Expression~\eqref{eq:shift2} appears to be pretty convenient in order to calculate the radial shift of the centroid, among two pairs of spin supplementary conditions, TD-MP and TD-OKS formalisms, in terms of the power series expansion method. In the following pages the above corrections will be imposed on the results of Sec.~\ref{sec:comp} for generic CEOs. Namely, we show how the convergence between the SSCs in terms of the orbital frequencies is affected, when we impose linear and quadratic corrections in the position of the centroid in terms of $\delta r$ and what happens when we take also into account the alteration of the value of the test body's spin  due to the change of the SSC.

\subsection{Linear Corrections}\label{sec:linCor}

In our first attempt to address the power series divergence between the SSCs demonstrated in Table~\ref{tab:CEOFrp}, we assume a linear correction of the position:
\begin{align}
  \tilde{r}=r+\delta r,  
\end{align}
with $\delta r$ given by Eq.~\eqref{eq:shift2}, but we do not correct the measure of the spin leaving $\tilde{\sigma}=\sigma$. The computation of the centroid's radial shift might appear complicated, since the RHS of Eq. \eqref{eq:shift2} pertains to $\delta r$, through the four-momentum $\tilde{p}_\mu$ and the dynamical rest mass $\tilde{\mu}$. By expanding Eq.~\eqref{eq:shift2} in the linear approximation in terms of a $\delta r$, we conclude that:
\begin{widetext}
\begin{align} \label{eq:deltar}
    \delta r =\frac{p_t S^{tr}+p_\phi S^{\phi r}}{\mu^2}
    +\frac{\delta r}{\mu^2}\biggl\{\frac{p_t g_{tt,r} S^{tr}}{g_{tt}}+\frac{p_\phi g_{\phi\phi ,r} S^{\phi r}}{g_{\phi\phi}}+\biggl(\frac{p_t S^{tr}+p_\phi S^{\phi r}}{\mu^2}\biggr)\biggl[g_{tt,r}\biggl(\frac{p_t}{g_{tt}}\biggr)^2+g_{\phi\phi ,r}\biggl(\frac{p_\phi}{g_{\phi\phi}}\biggr)^2\biggr]\biggr\}+\mathcal{O}(\delta r^2).
\end{align}
\end{widetext}

 By neglecting the $\delta r$-term on the RHS of Eq.~\eqref{eq:deltar}, one can derive a ``zero'' order approximation for the value of the radial shift. Note that, for the TD-MP pair, the various components of the four-momentum are given by relations \eqref{eq:MP1}, \eqref{eq:MP2}, with $p_t=g_{tt}p^t$ and $p_\phi=g_{\phi\phi}p^\phi$, while $S^{tr}$ and $S^{\phi r}$ are determined via Eqs.~\eqref{eq:s1}, \eqref{eq:s2}, with $V^\nu=u^\nu$. On the other hand, for the comparison of the TD-OKS pair, $p_t$ and $p_\phi$ in Eq.~\eqref{eq:deltar} have to be replaced by $p_t=mg_{tt}u^t$ as well as $p_\phi=mg_{\phi\phi}u^\phi$, whereas $S^{tr}$ and $S^{\phi r}$ are extracted, once again, from Eqs.~\eqref{eq:s1}, \eqref{eq:s2}, with $V^t$ and $V^\phi$ expressed through Eqs.~\eqref{eq:vt} and \eqref{eq:OKS1}, respectively. In both cases, the relation $u^\phi=\Omega u^t$ and Eq.~\eqref{eq:vel_t} have been taken into account, with $\Omega$ referring to the MP or the OKS SSC, correspondingly. \par      
 Once $\delta r$ has been evaluated, it shall be implemented in the first column of Table \ref{tab:CEOFrp}, in order to explicate the discrepancies in the orbital frequency. As we will see in the Tables~\ref{tab:CEOcorrP},~\ref{tab:CEOcorrOKS}, the introduction of the centroid's shift fixes the difference in the $\mathcal{O}(\sigma^3)$-term for the TD-MP pair of SSCs, and removes the disagreement in the $\mathcal{O}(\sigma^2)$-term between the TD and the OKS SSCs as well. In order to increase the accuracy of the proposed method we should also  take into account the $\delta r$-term in the RHS of Eq.~\eqref{eq:deltar}. By substituting the previous, zeroth order approximation for $\delta r$, the corrected orbital frequencies related to the TD SSC after the shift read:
 \begin{equation} \label{eq:orossigma4}
     \mathcal{O}(\sigma^4)=\frac{3(1453\hat{r}^2-6702\hat{r}+7605)}{128(\hat{r}-3)^2\sqrt{\hat{r}^{15}}},  
 \end{equation}
for the TD-MP pair of SSCs and:
\begin{equation} \label{eq:orossigma3}
    \mathcal{O}(\sigma^3)=\frac{6(8\hat{r}^2-48\hat{r}+63)}{4(\hat{r}-3)^2\hat{r}^6},
\end{equation}
 for the TD-OKS pair. The latter fact is a first indication that, we cannot infinitely improve the power series convergence, even when higher order corrections are imposed by assigning the same $\sigma$-value in all cases.    
 \begin{table}[h] 
\centering
\large
   \renewcommand{\arraystretch}{1.4}
  
   \begin{tabular}{ |c| c| c|}
\hline

$ \hat{\Omega}_n $ & TD SSC & MP SSC \\   \hline

$\mathcal{O}(\sigma^0)$ & $\frac{1}{\sqrt{\hat{r}^3}}$& $\frac{1}{\sqrt{\hat{r}^3}}$ \\ \hline
$\mathcal{O}(\sigma^1)$ &$-\frac{3}{2\hat{r}^3}$ & $-\frac{3}{2\hat{r}^3}$ \\ \hline
$\mathcal{O}(\sigma^2)$&$\frac{21}{8\sqrt{\hat{r}^9}}$& $\frac{21}{8\sqrt{\hat{r}^9}}$\\ \hline
$\mathcal{O}(\sigma^3)$&$-\frac{3(4\hat{r}-9)}{2(\hat{r}-3)\hat{r}^6}$& $-\frac{3(4\hat{r}-9)}{2(\hat{r}-3)\hat{r}^6}$\\ \hline
$\mathcal{O}(\sigma^4)$&$\frac{3(1069\hat{r}^2-4782\hat{r}+5301)}{128(\hat{r}-3)^2\sqrt{\hat{r}^{15}}}$& $\frac{3(685\hat{r}^2-2670\hat{r}+2421)}{128(\hat{r}-3)^2\sqrt{\hat{r}^{15}}}$ \\ \hline
     \end{tabular}
   \caption[caption]{The Table illustrates the power series expansion coefficients of the $\hat{\Omega}_+$ for the TD and MP SSCs, with $\tilde{r}\neq r$.}
   \label{tab:CEOcorrP}
\end{table}

\begin{table*}[ht] 
\centering
   \renewcommand{\arraystretch}{1.4}
  \large
   \begin{tabular}{ |c| c| c|}
\hline

$ \hat{\Omega}_n $ & TD SSC & OKS SSC \\   \hline

$\mathcal{O}(\sigma^0)$ & $\frac{1}{\sqrt{\hat{r}^3}}$& $\frac{1}{\sqrt{\hat{r}^3}}$ \\ \hline
$\mathcal{O}(\sigma^1)$ &$-\frac{3}{2\hat{r}^3}$ & $-\frac{3}{2\hat{r}^3}$ \\ \hline
$\mathcal{O}(\sigma^2)$&$-\frac{3(5\hat{r}-3)}{8(\hat{r}-3)\sqrt{\hat{r}^9}}$& $-\frac{3(5\hat{r}-3)}{8(\hat{r}-3)\sqrt{\hat{r}^9}}$\\ \hline
$\mathcal{O}(\sigma^3)$&$\frac{3(2\hat{r}^2-18\hat{r}+27)}{4(\hat{r}-3)^2\hat{r}^6}$& $-\frac{3(8\hat{r}^2-21\hat{r}+18)}{4(\hat{r}-3)^2\hat{r}^6}$\\ \hline
$\mathcal{O}(\sigma^4)$&$-\frac{3(971\hat{r}^3-5883\hat{r}^2+14409\hat{r}-13041)}{128(\hat{r}-3)^3\sqrt{\hat{r}^{15}}}$& $-\frac{3(971\hat{r}^3-3915\hat{r}^2+5625\hat{r}-2673)}{128(\hat{r}-3)^3\sqrt{\hat{r}^{15}}}$ \\ \hline
\end{tabular}
   \caption[caption]{The table shows the power series expansion coefficients of $\hat{\Omega}_+$  for the TD and the OKS SSCs, for CEOs at radii $\tilde{r}$ and $r$ respectively, when $\tilde{r}\neq r$.}
   \label{tab:CEOcorrOKS}
\end{table*}

\subsection{Spin measure correction}

As the centroid is shifted and measured with respect to another four-vector, the spin tensor changes as prescribed by Eq.~\eqref{eq:shift1}. Hence, we expect the spin measure to change as well; up to this point we have assumed $\tilde{\sigma}=\sigma$. This section investigates what are the consequences of the spin measure correction on the power series expansion of $\Omega_+$, that we have performed in the previous sections. Recalling that for CEOs the only non-vanishing components of the spin tensor are the $S^{tr}=- S^{rt}$ and $S^{r\phi}=- S^{\phi r}$, the spin measure~\eqref{eq:spin_m}, expanded in terms of the radial shift, reads:
\begin{align} \label{eq:spinmeasure}
  &\tilde{S}^2=S^2+\delta r \biggl\{g_{rr}\biggl[g_{\phi\phi,r}\biggl(S^{r\phi}\biggr)^2+g_{tt,r}\biggl(S^{tr}\biggr)^2 \notag\\
  &+2\biggl(p_t S^{tr}-p_\phi S^{r\phi}\biggr)\biggr]+\frac{S^2 g_{rr,r}}{g_{rr}}\biggr\}+\mathcal{O}\;\left(\delta r^2\right).  
\end{align}
Note that, the derivation of this relation follows after the assumption that both centroids move on CEOs and we have used Eq.~\eqref{eq:shift1} for each of the non-vanishing components of the TD SSC. This equation helps us to correlate the dimensionless spin, among two pairs of SSCs. This process however, is greatly SSC-dependent and as a result, it is examined in the two following subsections.
 
 \subsubsection{TD-MP relation}
 In order to obtain a correlation between $\tilde{\sigma}$ and $\sigma$, concerning the transition from the MP to the TD centroid, both sides of Eq.~\eqref{eq:spinmeasure} shall be divided by $\tilde{\mu}^2 M^2$. The inverse square of the dynamical rest mass in the linear in $\delta r$ approximation reduces to:
 \begin{equation} \label{eq:dmassTD2oSSC}
  \frac{1}{\tilde{\mu}^2} =\frac{1}{\mu^2}\biggl\{1+\frac{\delta r}{\mu^2}\biggl[g_{tt,r}\biggl(\frac{p_t}{g_{tt}}\biggr)^2+g_{\phi\phi,r}\biggl(\frac{p_\phi}{g_{\phi\phi}}\biggr)^2\biggr]\biggr\} 
  +\mathcal{O}\left(\delta r^2\right). 
 \end{equation}
 Recall that, under MP SSC, the dimensionless spin is written as $\sigma=\frac{S}{mM}$. Hence, a link between $m$ and $\mu$ is essential for deriving the dimensionless spin $\tilde{\sigma}$ measured in the TD reference frame, in terms of its counterpart $\sigma$, measured in the MP frame. Such a connection is provided \cite{Costa18} by:
 \begin{equation}
     \mu^2=m^2+\frac{S^{\alpha\kappa}S_{\kappa\beta}p^\beta p_\alpha}{S^2}.
 \end{equation}
 Hence, one has to express $\tilde{\mu}$ as a function of $\mu$ and relate them to $m$ in the MP frame. The aforementioned process yields up to $\mathcal{O}\left(\delta r\right)$ the following relation:
 \begin{widetext}
 \begin{equation}
  \frac{1}{\tilde{\mu}^2}=\frac{S^2}{m^2 S^2-g_{rr}(p_t S^{tr}-p_\phi S^{r\phi})^2}\biggl\{1+\frac{ S^2 \delta r}{m^2 S^2-g_{rr}(p_t S^{tr}-p_\phi S^{r\phi})^2}\biggl[g_{tt,r}\biggl(\frac{p_t}{g_{tt}}\biggr)^2+g_{\phi\phi,r}\biggl(\frac{p_\phi}{g_{\phi\phi}}\biggr)^2\biggr]\biggr\},
 \end{equation}
\end{widetext}
 which then leads to the spin measure correction: 
 \begin{widetext}
  \begin{align} \label{eq:spin_MP2TD}
     &\tilde{\sigma}^2 =\frac{\sigma^2}{\sigma^2-g_{rr}(p_t \sigma^{tr}-p_\phi \sigma^{r\phi})^2/m^2}\Biggl\{\sigma^2+\delta r\biggl\{ \frac{\sigma^4}{\sigma^2-g_{rr}(p_t \sigma^{tr}-p_\phi \sigma^{r\phi})^2/m^2} \biggl[g_{tt,r}\biggl(\frac{p_t}{m~g_{tt}}\biggr)^2\notag
    +g_{\phi\phi,r}\biggl(\frac{p_\phi}{m~g_{\phi\phi}}\biggr)^2\biggr] \nonumber \\
    &+g_{rr}\biggl[g_{\phi\phi,r}\biggl(\sigma^{r\phi}\biggr)^2+g_{tt,r}\biggl(\sigma^{tr}\biggr)^2
    +\frac{2}{m~M}\biggl(p_t \sigma^{tr}-p_\phi \sigma^{r\phi}\biggr)\biggr]+\frac{\sigma^2 g_{rr,r}}{g_{rr}}\biggr\}\Biggr\},
    \end{align}
 \end{widetext}
where we have introduced the normalized\footnote{The components of the normalized spin tensor  $\sigma^{\mu\nu}$ are not necessarily dimensionless. More specifically, $[\sigma^{r\phi}]=[M]^{-1}$, while $\sigma^{tr}$ is actually dimensionless.} spin tensor $\sigma^{\kappa\nu}=S^{\kappa\nu}/(mM) $ to make the spin measures relation~\eqref{eq:spin_MP2TD} more compact. 

The above lengthy expression~\eqref{eq:spin_MP2TD}  yields a much more compact form for a CEO in the Schwarzschild case, and especially after  the implementation of a power series expansion with respect to $\sigma$, it reduces to:
\begin{equation} \label{eq:sTDMP}
    \tilde{\sigma}=\sigma \biggl[1+\frac{3(\hat{r}-2)\sigma^3}{(\hat{r}-3)\hat{r}^{\frac{9}{2}}}\biggr]+\mathcal{O}\left(\sigma^5\right).
\end{equation}
From Eq.~\eqref{eq:sTDMP} it is obvious that, further corrections on the test body's orbital frequency only occur at quartic, or higher order terms. Namely, the power series with respect to $\sigma$ in the first column of Table \ref{tab:CEOcorrP} has to be substituted by $\tilde{\sigma}$. Whereas the described procedure does not affect the  lower (up to $\sigma^3$) order contributions, the power series convergence is not improved either, since after rewritting $\tilde{\sigma}$ as a function of $\sigma$:
\begin{equation}
    \mathcal{O}(\sigma^4)=\frac{3(877\hat{r}^2-3822\hat{r}+4149)}{128(\hat{r}-3)^2\sqrt{\hat{r}^{15}}}.
\end{equation}
As we will see in the next section, this appears to be a general trend, which is also present in the OKS to TD case.

\subsubsection{TD-OKS relation}

The derivation of a relation similar to Eq. \eqref{eq:sTDMP} is fairly simpler, under the transition from OKS to TD SSC. The reason behind this fact is the constancy of the dynamical rest mass of the test body, when OKS is imposed. Once again, one starts from Eq.~\eqref{eq:spinmeasure}, by dividing both sides by $\tilde{\mu}^2 M^2$ and using Eq.~\eqref{eq:dmassTD2oSSC} on the RHS, that straightforwardly gives: 
\begin{widetext}
\begin{equation} \label{eq:dspinOKS2TD}
    \tilde{\sigma}^2 =\sigma^2+\delta r\biggl\{g_{rr}\biggl[g_{\phi\phi,r}\biggl(\sigma^{r\phi}\biggr)^2+g_{tt,r}\biggl(\sigma^{tr}\biggr)^2     +\frac{2}{\mu M} \biggl(p_t \sigma^{tr}-p_\phi \sigma^{r\phi}\biggr) \biggr]+\sigma^2\biggl[\frac{g_{rr,r}}{g_{rr}}+g_{tt,r}\biggl(\frac{p_t}{\mu~g_{tt}}\biggr)^2 
 +g_{\phi\phi,r}\biggl(\frac{p_\phi}{\mu~g_{\phi\phi}}\biggr)^2 \biggr]\biggr\},
 \end{equation}
\end{widetext}
with the definition of the normalized spin tensor $\sigma^{\kappa\nu}$, equivalently to the TD-MP case. 

For Schwarzschild spacetime, the expression~\eqref{eq:dspinOKS2TD} reduces to:
\begin{equation} \label{eq:sTDOKS}
    \tilde{\sigma}=\sigma\biggl\{1+\frac{3(\hat{r}-2)\sigma^2}{[\hat{r}(\hat{r}-3)]^2}\biggr\}+\mathcal{O}\left(\sigma^4\right),
\end{equation}
by taking advantage of the power series expansion method. In agreement with the former set of spin supplementary conditions, Eq. \eqref{eq:sTDOKS} fails to further improve the orbital frequency's power series convergence. More specifically, this  expression imposes modifications to the term proportional to $\sigma^3$, in the first column of Table \ref{tab:CEOcorrOKS}, namely:
\begin{equation}
    \mathcal{O}(\sigma^3)=-\frac{6(\hat{r}^2+12\hat{r}-27)}{4(\hat{r}-3)^2\hat{r}^6},
\end{equation}
which still remains  different from the corresponding term, produced by the OKS SSC.

\subsection{Quadratic Corrections}

If we don't keep just the lower order terms in the aforementioned analysis, we could proceed one step beyond, by including non-linear terms to the shift from a SSC centroid to the TD centroid. The general framework is for example provided in \cite{Costa18}. In a quadratic in shift correction the worldline $\tilde{z}^\nu$ is related to $z^\nu$ as follows:
\begin{equation} \label{eq:cost}
    \tilde{z}^\nu=z^\nu+\delta z^\nu-\frac{1}{2}\Gamma ^\nu_{\kappa\pi}\delta z^\kappa\delta z^\pi.
\end{equation}
For CEOs in Schwarzschild spacetime Eq.~\eqref{eq:cost} reduces to:
\begin{equation} \label{eq:quadshift}
    \tilde{r}=r+\delta r+\frac{M \delta r^2}{2r(r-2M)}+\mathcal{O}\left(\delta r^3\right).
\end{equation}

Again, Eq.~\eqref{eq:shift2} is the required relation, in order to develop a procedure, similar to that produced for the linear corrections in Sec.~\ref{sec:linCor}. By multiplying both sides of Eq.~\eqref{eq:shift2} by $\tilde{\mu}^2=-\tilde{g}_{\kappa\sigma}p^\kappa p^\sigma$, while raising the indices of the four-momentum on the numerator, at the same time, implies that:
\begin{equation}
   \tilde{g}_{tt}p^t\biggl(\delta r p^t -S^{rt}\biggr)=\tilde{g}_{\phi\phi}p^\phi \biggl(S^{r\phi}-\delta r p^\phi \biggr).
\end{equation}
The non vanishing components of the metric tensor, measured in the TD frame, can be Taylor expanded in terms of the radius $\tilde{r}$ of Eq.~\eqref{eq:quadshift}. This leads to a quadratic equation with respect to the radial shift, that is:
\begin{widetext}
\begin{align}
 &\biggl\{2r\biggl(r-2M\biggr)\biggl[M\biggl(p^t\biggr)^2-r^3\biggl(p^\phi \biggr)^2\biggr]+M\biggl(2r-5M\biggr)p^t S^{rt}+r^3\biggl(r-M\biggr)p^\phi S^{r\phi}\biggr\}\delta r^2+r\biggl(r-2M\biggr)\nonumber\\
 &\times \biggl\{r\biggl[\biggl(r-2M\biggr)\biggl(p^t\biggr)^2-r^3\biggl(p^\phi\biggr)^2\biggr]-2\biggl[M p^t S^{rt}-r^3 p^\phi S^{r\phi}\biggr]\biggr\}\delta r +r^2\biggl(r-2M\biggr)\biggl[r^3 p^\phi S^{r\phi}-\biggl(r-2M\biggr)p^t S^{rt}\biggr]=0, 
\end{align}
\end{widetext}
which in general has two distinct roots. One of them is completely unnatural, since it approaches infinity in the geodesic limit $S\to 0$. On the other hand, the physically acceptable solution:
\begin{widetext}
\begin{equation}
    \delta r=\frac{r\biggl[\biggl(2M-r\biggr)\biggl(p^t\biggr)^2+r^3\biggl(p^\phi\biggr)^2-\biggl(2M-r\biggr)\biggl(p^t\biggr)^2-r^3\biggl(p^\phi\biggr)^2\biggr]}{4\biggl[M\biggl(p^t\biggr)^2-r^3\biggl(p^\phi \biggr)^2\biggr]},
\end{equation}
\end{widetext}
has a $\frac{0}{0}$ indeterminacy, in the geodesic limit, which can be overcome by applying L'H\^{o}pital's rule, in order to obtain:
\begin{equation*}
    \lim_{S \to 0} \delta r \propto\biggl[ M\biggl(p^t\biggr)^2-r^3\biggl(p^\phi\biggr)^2\biggr]=0\, ,
\end{equation*}
as expected.

Applying the quadratic corrections of the centroid on the first column of Table~\ref{tab:CEOFrp} (substitution of $r$ by $\tilde{r}$) without proceeding to the spin measure correction, slightly alters the difference in the $\mathcal{O}\left(\sigma^4\right)$-term of the TD-MP pair as well as in the $\mathcal{O}\left(\sigma^3\right)$-term of the TD-OKS pair, but does not improve the results more than those shown in Tables~\ref{tab:CEOcorrP} and \ref{tab:CEOcorrOKS}, respectively. More precisely, the aforementioned terms are equal to those presented in Eqs.~\eqref{eq:orossigma4} and \eqref{eq:orossigma3}, correspondingly. The convergence does not improve even if we take into account the spin measure correction, then the TD-MP pair gives:
\begin{equation}
    \mathcal{O}(\sigma^4)=\frac{3(1261\hat{r}^2-5742\hat{r}+6453)}{128(\hat{r}-3)^2\sqrt{\hat{r}^{15}}},
\end{equation}
and the  TD-OKS pair yields:
\begin{equation}
    \mathcal{O}(\sigma^3)=\frac{6(5\hat{r}^2-42\hat{r}+63)}{4(\hat{r}-3)^2\hat{r}^6}.
\end{equation}

It seems that for higher order corrections than those employed for Tables~\ref{tab:CEOcorrP} and \ref{tab:CEOcorrOKS}, the power series convergence starts to diminish. We have strong indications that this behaviour has a theoretical background, the nature of which, will be discussed in Sec.~\ref{sec:concl}. A subsequent consequence of the aforementioned observation is that higher order corrections can be neglected , since the procedure appears to have already reached its peak convergence. For example, the convergence would not improve if we took into account $\mathcal{O}(\delta r^2)$ terms in Eq.~\eqref{eq:spinmeasure}.

\section{Conclusions} \label{sec:concl}

The equivalence of Mathisson-Papapetrou-Dixon equations under various spin supplementary conditions has ignited controversy among the community, over the years. The motivation of the present study is to investigate the discrepancies in the spinning test body's orbital frequency along circular equatorial orbits, which are often studied in the literature \cite{Harms16,Tanaka:1996ht,Suzuki:1997by,Hackmann14,Costa18,Tod76}. To achieve this we introduced an analytical algorithm giving the frequencies for any SSC on an arbitrary, stationary, axisymmetric spacetime, with reflection symmetry and established a power series analysis with respect to spin magnitude, in order to examine the differences between the SSCs in a more thorough and convenient  quantitative way.

First and foremost, even without the introduction of centroid corrections, it is apparent that the Tulczyjew-Dixon choice is more compatible with the Mathisson-Pirani SSC, than the Ohashi-Kyrian-Semer\'{a}k formalism. The latter fact follows from the orbital frequency convergence, presented in Table~\ref{tab:CEOFrp}, in which we see that in the TD-MP comparison the series converge up to the quadratic term in spin, while for the TD-OKS comparison the convergence is just linear in spin. Looking at the innermost stable circular orbits, the convergence is improved by one order in spin for both comparisons, as demonstrated in Table~\ref{tab:ISCOFrp}. This improvement in the series convergence was also achieved, when the linear correction in the radial position of the orbits was taken into account. However, by applying shift and spin corrections simultaneously, we were not able to further improve the power series convergence. The same negative result was obtained for quadratic corrections in the radial shift of the centroids. As a result, it is safe to argue that, all three examined SSCs converge with each other up to quadratic spin terms in the case of circular equatorial orbits around a Schwarzschild black hole. 

It is crucial to notice at this point that, the origin of the observed disagreements between SSCs, is deeply nested inside the expression of the radial shift in Eq.~\eqref{eq:shift2}. Furthermore, the fact that additional improvements to the power series of the frequencies with respect to $\sigma$ were not fruitful, pertains to an interesting theoretical reason. More precisely, the circular orbits of the spinning test body's centroid could potentially degenerate into plain, but not circular, equatorial orbits, under the alteration of SSC reference frame. Namely, since $p^r=0$ for a CEO under MP or OKS SSC, then by shifting the centroid to TD SSC at least initially $\tilde{p}^r=0$, which might just be related  to a turning point of the corresponding trajectory. This can be even caused from the fact that the transition rules are followed up to certain order in the radial shift and spin expansion. Hence, the failure to shift from an arbitrary CEO under a certain SSC, to another CEO, governed by a different SSC, is responsible for the divergence of the orbital frequency power series. This interpretation might be even consistent with the numerical findings of \cite{Costa18} examining shifts between non-helical to helical MP SSC setups. However, the most probable interpretation is that the pole-dipole approximation breaks down in curved spacetimes, which implies that the shift between SSCs is not simply a gauge transformation as in the flat spacetime \cite{Costa18}.

To exclude one of the above interpretations, let us focus on the ISCO results. It is interesting to note that the order of convergence in the series expansions of the frequencies $\Omega$ for ISCO (Table~\ref{tab:ISCOFrp}) between the SSCs is of the same order as the order of convergence for arbitrary CEOs after appropriate corrections have been applied (Tables~\ref{tab:CEOcorrP},\ref{tab:CEOcorrOKS}). This implies that only by investigating the ISCO case one could tell the maximal possible convergence between the CEOs of different SSCs. This fact  eliminates the aforementioned turning point interpretation. Why ISCO is so special CEO? Probably it has to do with the fact that ISCO is a stability limit and hence, it should  not be possible to apply any correction to this limit due to a centroid shift. If we have ISCO for one centroid, it should be ISCO for all the centroids of the same physical body. This does not mean that all the centroids would lie at the same radius, since changing a SSC means that we change the reference point in the body. Actually, note that the convergence of the radii in Table~\ref{tab:ISCOr} is smaller than the convergence of the frequencies in Table~\ref{tab:ISCOFrp} by one order in the spin expansion. Thus, centroids lying at radii that are appropriately different at some order tend to lead to frequencies that are equal up to higher order terms of the expansion. In conclusion, if we take into account the shift of centroids from one SSC to another, we can bring the rotation frequencies of CEOs into better agreement in terms of a power series expansion with respect to the magnitude of spin, but this process terminates at some order, where the pole-dipole approximation breaks down, and one has to take into account higher order multipoles as well.

\begin{acknowledgments}
The authors would like to acknowledge networking support by the GWverse COST Action CA16104, ``Black holes, gravitational waves and fundamental physics''. G.L.G. would also like to express gratitude for the hospitality of the Department of Physics at the University of Athens. G.L.G. has been supported by the fellowship Lumina Quaeruntur No. LQ100032102 of the Czech Academy of Sciences.
\end{acknowledgments}

\appendix
\section{Further Analysis On SAR Spacetimes} \label{sec:app1}

This section of the paper presents long expressions not shown in Sec.~\ref{sec:CEOs} in order to make the text more readable. The structure of Appendix~\ref{sec:app1} is organized as follows. The first part covers the details regarding the quadratic equation, satisfied by the test body's orbital frequency, under the Tulczyjew-Dixon condition. Additionally, in the second part we discuss the considerably more complex case of the quartic equation in $\Omega$, which is derived for the Mathisson-Pirani SSC, for a generic SAR spacetime.

\subsection{Tulczyjew-Dixon SSC} 

The coefficients $\rho_1$, $\rho_2$, $\rho_3$ of Eq.~\eqref{eq:2vathmia} read:
\begin{widetext}
\begin{align*}
     &\rho_1=-g\mu^2\Gamma^r_{\phi\phi}+g\mu S \sqrt{-\frac{ g_{\theta\theta}}{g}}\biggl[-g_{tt}\Gamma^t_{\phi r}\Gamma^r_{\phi\phi}+g_{\phi\phi}\biggl(\Gamma^r_{t\phi}\Gamma^\phi_{\phi r}+R^r_{\phi t r}\biggr)+g_{t\phi}\biggl(\Gamma^t_{\phi r}\Gamma^r_{t\phi}-\Gamma^r_{\phi\phi}\Gamma^\phi_{\phi r}+R^r_{\phi r \phi}\biggr)\biggr]\\
     &+g_{\theta\theta}S^2\biggl(g_{t\phi}^2-g_{tt}g_{\phi\phi}\biggr)\biggl(\Gamma^t_{\phi r}R^r_{\phi t r}-\Gamma^\phi_{\phi r}R^r_{\phi r \phi}\biggr),\\
     &\rho_2=-2g\mu^2\Gamma^r_{t\phi}+g\mu S \sqrt{-\frac{ g_{\theta\theta}}{g}}\biggl[g_{t\phi}\biggl(\Gamma^t_{\phi r}\Gamma^r_{tt}+\Gamma^t_{tr}\Gamma^r_{t\phi}-\Gamma^r_{\phi\phi}\Gamma^\phi_{tr}-\Gamma^r_{t\phi}\Gamma^\phi_{\phi r}\biggr)+g_{\phi\phi}\biggl(\Gamma^r_{t\phi}\Gamma^\phi_{tr}+\Gamma^r_{tt}\Gamma^\phi_{\phi r}+R^r_{ttr}\biggr)\\
     &-g_{tt}\biggl(\Gamma^t_{\phi r}\Gamma^r_{t\phi}+\Gamma^t_{tr}\Gamma^r_{\phi\phi}-R^r_{\phi r \phi}\biggr)\biggr]+g_{\theta\theta}S^2\biggl(g_{t\phi}^2-g_{tt}g_{\phi\phi}\biggr)\biggl(\Gamma^t_{\phi r}R^r_{ttr}+\Gamma^t_{tr}R^r_{\phi t r}+\Gamma^\phi_{\phi r}R^r_{\phi t r}-\Gamma^\phi_{tr}R^r_{\phi r \phi}\biggr),\\
     &\rho_3=-g\mu^2\Gamma^r_{tt}+g\mu S \sqrt{-\frac{ g_{\theta\theta}}{g}}\biggl[g_{\phi\phi}\Gamma^r_{tt}\Gamma^\phi_{tr}+g_{t\phi}\biggl(\Gamma^t_{tr}\Gamma^r_{tt}-\Gamma^r_{t\phi}\Gamma^\phi_{tr}+R^r_{ttr}\biggr)-g_{tt}\biggl(\Gamma^t_{tr}\Gamma^r_{t\phi}+R^r_{\phi tr}\biggr)\biggr]\\
     &+g_{\theta\theta}S^2\biggl(g_{t\phi}^2-g_{tt}g_{\phi\phi}\biggr)\biggl(\Gamma^t_{tr}R^r_{ttr}+\Gamma^\phi_{tr}R^r_{\phi tr}\biggr),
\end{align*}
\end{widetext}
and hold for every SAR spacetime. It is worth noting that the discriminant $\rho_2^2-4\rho_1\rho_3$ has to be positive in order to obtain CEOs.

\subsection{Mathisson-Pirani SSC}
The polynomial coefficients $\xi_i$ of Eq.~\eqref{eq:4vathmia}  are given by the following analytic expressions:
\begin{widetext}
\begin{align*}
    &\xi_1=m g_{\phi\phi}\Gamma^r_{\phi\phi}+S\sqrt{-\frac{ g_{\theta\theta}}{g}}\biggl[g_{t\phi}^2\Gamma^t_{\phi r}\Gamma^r_{\phi\phi}+g_{\phi\phi}^2\biggl(R^r_{tr\phi}-\Gamma^r_{t\phi}\Gamma^\phi_{\phi r}\biggr)-g_{t\phi}g_{\phi\phi}\biggl(\Gamma^t_{\phi r}\Gamma^r_{t\phi}-\Gamma^r_{\phi\phi}\Gamma^\phi_{\phi r}+R^r_{\phi r \phi}\biggr)\biggr],\\
    &\xi_2=2m\biggl(g_{\phi\phi}\Gamma^r_{t\phi}+g_{t\phi}\Gamma^r_{\phi\phi}\biggr)+S \sqrt{-\frac{ g_{\theta\theta}}{g}}\biggl\{-g_{t\phi}\biggl[-2g_{tt}\Gamma^t_{\phi r}\Gamma^r_{\phi\phi}+g_{\phi\phi}\biggl(\Gamma^t_{\phi r}\Gamma^r_{tt}+\Gamma^t_{tr}\Gamma^r_{t\phi}-\Gamma^r_{\phi\phi}\Gamma^\phi_{tr}+\Gamma^r_{t\phi}\Gamma^\phi_{\phi r}-2R^r_{tr\phi}\biggr)\biggr]\\
    &+g_{t\phi}^2\biggl(\Gamma^t_{tr}\Gamma^r_{\phi\phi}+\Gamma^r_{\phi\phi}\Gamma^\phi_{\phi r}-2R^r_{\phi r \phi}\biggr)-g_{\phi\phi}\biggl[g_{\phi\phi}\biggl(\Gamma^r_{t\phi}\Gamma^\phi_{tr}+\Gamma^r_{tt}\Gamma^\phi_{\phi r}+R^r_{ttr}\biggr)+g_{tt}\biggl(\Gamma^t_{\phi r}\Gamma^r_{t\phi}-\Gamma^r_{\phi\phi}\Gamma^\phi_{\phi r}+R^r_{\phi r \phi}\biggr)\biggr]\biggr\},\\
    &\xi_3=m\biggl(g_{\phi\phi}\Gamma^r_{tt}+4g_{t\phi}\Gamma^r_{t\phi}+g_{tt}\Gamma^r_{\phi\phi}\biggr)\\&-S \sqrt{-\frac{ g_{\theta\theta}}{g}}\biggl\{-g_{tt}^2\Gamma^t_{\phi r}\Gamma^r_{\phi\phi}+g_{\phi\phi}^2\Gamma^r_{tt}\Gamma^\phi_{tr}+g_{t\phi}^2\biggl(\Gamma^t_{\phi r}\Gamma^r_{tt}-\Gamma^r_{\phi\phi}\Gamma^\phi_{tr}\biggr)+g_{tt}g_{\phi\phi}\biggl(\Gamma^t_{\phi r}\Gamma^r_{tt}+\Gamma^t_{tr}\Gamma^r_{t\phi}-\Gamma^r_{\phi\phi}\Gamma^\phi_{tr}-\Gamma^r_{t\phi}\Gamma^\phi_{\phi r}\biggr)\\
    &+g_{t\phi}\biggl[g_{\phi\phi}\biggr(\Gamma^t_{tr}\Gamma^r_{tt}+\Gamma^r_{t\phi}\Gamma^\phi_{tr}+2\Gamma^r_{tt}\Gamma^\phi_{\phi r}+3R^r_{ttr}\biggr)-g_{tt}\biggl(\Gamma^t_{\phi r}\Gamma^r_{t\phi}+2\Gamma^t_{tr}\Gamma^r_{\phi\phi}+\Gamma^r_{\phi\phi}\Gamma^\phi_{\phi r}-3R^r_{\phi r \phi}\biggr)\biggr]\biggr\},\\
    &\xi_4=2m\biggl(g_{t\phi}\Gamma^r_{tt}+g_{tt}\Gamma^r_{t\phi}\biggr)-S \sqrt{-\frac{ g_{\theta\theta}}{g}}\biggl\{g_{t\phi}^2\biggl(\Gamma^t_{tr}\Gamma^r_{tt}+\Gamma^r_{tt}\Gamma^\phi_{\phi r}+2R^r_{ttr}\biggr)+g_{t\phi}\biggl[2g_{\phi\phi}\Gamma^r_{tt}\Gamma^\phi_{tr}+g_{tt}\biggl(\Gamma^t_{\phi r}\Gamma^r_{tt}-\Gamma^t_{tr}\Gamma^r_{t\phi}\\
    &-\Gamma^r_{\phi\phi}\Gamma^\phi_{tr}-\Gamma^r_{t\phi}\Gamma^\phi_{\phi r}+2R^r_{tr\phi}\biggr)\biggr]+g_{tt}\biggl[g_{\phi\phi}\biggl(\Gamma^t_{tr}\Gamma^r_{tt}-\Gamma^r_{t\phi}\Gamma^\phi_{tr}+R^r_{ttr}\biggr)-g_{tt}\biggl(\Gamma^t_{\phi r}\Gamma^r_{t\phi}+\Gamma^t_{tr}\Gamma^r_{\phi\phi}-R^r_{\phi r \phi}\biggr)\biggr]\biggr\},\\
    &\xi_5=mg_{tt}\Gamma^r_{tt}-S \sqrt{-\frac{ g_{\theta\theta}}{g}}\biggl[g_{t\phi}^2\Gamma^r_{tt}\Gamma^\phi_{tr}-g_{tt}^2\biggl(\Gamma^t_{tr}\Gamma^r_{t\phi}-R^r_{tr\phi}\biggr)+g_{tt}g_{t\phi}\biggl(\Gamma^t_{tr}\Gamma^r_{tt}-\Gamma^r_{t\phi}\Gamma^\phi_{tr}+R^r_{ttr}\biggr)\biggr].
\end{align*}
\end{widetext}

\section{General Shifts} \label{sec:shifts}

This section summarizes the discussion of Sec.~\ref{sec:Corr} on centroid shifts and we show that the radial shift, constrained by the four-momentum, is one of a few viable options. We investigate the possibility of different radial or non radial shifts for circular equatorial orbits. We keep the symbolism of  Sec.~\ref{sec:Corr}, where the tilde symbol has been used to designate the quantities measured in the TD reference frame.    

\subsection{Momentum Shift Constraint in a non-radial direction}

Let us assume a non radial shift for the centroid on the equatorial plane, i.e. ($\delta r=\delta \theta=0$), along with the momentum shift constraint~\eqref{eq:shit_con}, which provides that:
\begin{equation} \label{eq:app21}
    \frac{\delta t}{\delta \phi}=-\frac{\tilde{p}_\phi}{\tilde{p}_t} \,.
\end{equation}
In this case, the transformation equation~\eqref{eq:shift1} of the spin tensor has only one pair of non trivial components: 
\begin{equation} \label{eq:app22}
    S^{t\phi}=p^\phi \delta t-p^t \delta \phi=-S^{\phi t} \, ,
\end{equation}
as $\tilde{S}^{t\phi}=0$. For a non-circular equatorial orbit it holds that:
\begin{align}
   S^{t\phi} = -S^{\phi t} = S~V_r \sqrt{-\frac{g_{\theta\theta}}{g}} \, . 
\end{align}
If we assume that $z^\nu$ also lies on a CEO, then the component $S^{t \phi}$ vanishes for both MP and OKS choices, since in both cases the radial component of the reference vector $V^\nu$ should be zero \cite{Harms16}. As a result Eq.~\eqref{eq:app22} reads:
\begin{equation} \label{eq:app23}
    \frac{\delta t}{\delta \phi}=\frac{p^t}{p^\phi},
\end{equation}
or by equating the RHS of Eqs.~\eqref{eq:app21} and \eqref{eq:app23}, one concludes that $\tilde{\mu}^2=0$, which cannot hold for a massive test body. Thus, there cannot be a non-radial shift of a MP/OKS centroid on a CEO to a TD centroid on a CEO, under the constraint~\eqref{eq:shit_con}.

\subsection{SSC Vector Shift Constraint}

Let us now impose that the centroid's shift obeys $V_\nu \delta z^\nu=0$, with $V^\nu$ being a generic  timelike four-vector, satisfying $V_\nu S^{\mu\nu}=0$. The contraction of Eq.~\eqref{eq:shift1} with the tensor quantity $\tilde{p}_\nu V_\mu$ yields:
\begin{equation} \label{eq:app24}
    \tilde{p}_\nu V_\kappa \delta z^\nu p^\kappa   =0 \, .
\end{equation}
Thus, Eq.~\eqref{eq:app24} straightforwardly implies that, either $\tilde{p}_\nu \delta z^\nu=0$, or $p^\kappa V_\kappa =0$. The former choice has been discussed in the previous section of Appendix \ref{sec:shifts} as well as in Sec.~\ref{sec:comp}, while the latter option is forbidden, due to the timelike character of the four-vectors $p^\nu$ and $V^\nu$. 

\subsection{A Timelike Vector To Constrain The Shift}

In the last part of Appendix \ref{sec:shifts} we use a timelike four-vector to restrict the shift, $W_\nu \delta z^\nu=0$, which does not satisfy any particular SSC, i.e. $W_\mu S^{\mu\nu}\neq0$.  Under this assumption, Eq.~\eqref{eq:shift1} leads to:
\begin{equation} \label{eq:app2last}
    \delta z^\nu=\frac{1}{\tilde{\mu}^2}\biggl(\tilde{p}_\mu S^{\mu\nu}+\frac{\tilde{p}_\sigma W_\lambda p^\nu S^{\lambda\sigma}}{W_\kappa p^\kappa }\biggr) \,.
\end{equation}
Since the orbit for the MP or the OKS frame lies on the equatorial plane, $p^\theta=0$, which implies from Eq.~\eqref{eq:app2last} that, $\delta\theta\propto \tilde{p}_\mu S^{\mu\theta}$. This suggests that $S^{\mu\theta}=0$ is a viable choice, if we opt having an equatorial orbit after the shift. Additionally, Eq. \eqref{eq:app2last} indicates that:
\begin{equation} \label{eq:app2telos}
    \tilde{p}_t S^{tr}+\tilde{p}_\phi S^{\phi r}=0 \,,
\end{equation}
under the hypothesis of a non radial shift ($\delta r=0)$, since $p^r=0$ is a common CEO feature, shared by all SSCs. The combination of Eqs.~\eqref{eq:app2last},~\eqref{eq:app2telos} along with the expressions $S^{t\phi}=0$ and $\tilde{p}_r=0$ gives $\delta t=\delta\phi=0$. Hence, we cannot shift from a circular equatorial orbit under TD SSC, to reach a CEO under another SSC, with a non radial shift. 

By introducing a timelike vector $W^\mu$, which could provide a non-radial shift between centroids lying on CEOs, would be an interesting choice in SAR spacetimes, since SAR metric elements do not depend on the coordinate time and the azimuthal angle. But, since we cannot achieve non-radial shift and since $W^\mu$ does not correspond to a specific SSC, such a choice is expected to introduce more undesirable complications than convenience.

\bibliographystyle{unsrt}
\bibliography{refs}

\end{document}